# Manuscript

# LiquiFab – Building with liquids in weightlessness

Erez Hochman[1,2], Aaron Sprecher[2,*], Kateryna Suzina[2], Amir Mann[3], Yuval Mihalovich[2], Valeri Frumkin[4], Moran Bercovici[1,5*]

[1] Faculty of Mechanical Engineering, Technion – Israel Institute of Technology, Haifa, Israel
[2] Faculty of Architecture, Technion – Israel Institute of Technology, Haifa, Israel
[3] Faculty of Computer Science, Technion – Israel Institute of Technology, Haifa, Israel
[4] Boston University, Boston, MA, USA
[5] Department of Materials, ETH Zürich, Zürich, Switzerland

\* Corresponding authors: asprecher@technion.ac.il, moran.bercovici@mat.ethz.ch

Existing digital manufacturing methods can be broadly divided into subtractive approaches, where material is removed from a bulk to reveal the desired form, and additive methods, in which material is introduced voxel-by-voxel to create an object.

We here show a fundamentally different method for the fabrication of three-dimensional objects that is neither subtractive nor additive. Instead of removal or layer-by-layer material deposition, in LiquiFab we shape an entire volume of liquid polymer by subjecting it to a set of geometrical constraints under conditions of weightlessness. The physics of liquid interfaces then drives the polymer to naturally adopt a configuration that minimizes its surface energy. On Earth, we achieve weightlessness through neutral buoyancy, and show that a small, well-defined set of boundary surfaces can be used to drive the liquid into a desired form that is then solidified. By sequentially applying this process, complex architectures can be assembled from successive liquid-formed elements.

Unlike additive manufacturing, where every point within the object must be individually visited by a print head or light field, LiquiFab forms the entire structure simultaneously. This makes the process highly scalable and opens the door to rapid manufacturing of large objects both on Earth and in space.

## Introduction

Form-finding is an approach in which the shape of a surface or structure is defined by a natural phenomenon that manifests itself in three dimensions (e.g., soap bubbles) or has a three-dimensional representation (e.g., energy level around an atom). The designer cannot directly alter the underlying physics, but can nonetheless affect the form by controlling a set of parameters or boundary conditions.[1]

One of the prominent examples of form finding dates back to the art nouveau architect Antoni Gaudí in the late 19th century, who suspended interconnected chains from fixed points at different distances and hung weights (sandbags) at various locations along the chains. Under gravity, the chains take a discontinuous catenary shape, which Gaudí replicated in his structures using standard construction techniques.[2] Starting in the early 1960s, architect Frei Otto explored the creation of surfaces by subjecting soap bubbles to various spatial boundary conditions.[3] Similar to Gaudi, his exploration was used only in the design process, and the actual structures were implemented using existing construction techniques of the time. Form-Finding can also occur digitally, as first demonstrated in the 1990s by architect Greg Lynn[4]. He predicted new forms using a 3D computational 'Blobiness' model originally developed by James Blinn[5] for graphically describing electron clouds in an atom.

Recently, Frumkin and Bercovici[6] and Elgarisi et al[7], developed a new fabrication method for optical components that is based on a volume of liquid polymer naturally finding its form when coming in contact with a bounding frame under weightlessness conditions. The designer cannot change the laws of physics governing the shape, but can modify the bounding frame and liquid volume to obtain a desired optical function. The authors showed the implementation of this method both in space[8] and on Earth, using neutral buoyancy conditions to counter gravity.

We here expand this approach into a more general form-finding fabrication method, which we term LiquiFab. LiquiFab creates three-dimensional objects by bringing a liquid volume into contact with a small set of surfaces purposely positioned in space. In contrast to 3D printing, where objects are built by following a toolpath (FDM)[9] or by sequentially solidifying layers (SLA or DLP)[10], LiquiFab is scale-invariant and produces the final object instantaneously once the injected liquid comes into contact with its boundaries. The processing time is only that of the injection and polymerization, which scale only weakly with volume. Both full and hollow structures, including ones with multiple partitions, can be formed, and in all cases the resulting surfaces exhibit sub-nanometric surface roughness without the need for post-processing. Importantly, in contrast to historical form-finding methods that were used exclusively to design complex morphologies that were then implemented using existing tools and materials, LiquiFab utilizes form-finding principles for the construction process itself.

More complex structures can be formed by sequentially applying LiquiFab; once an object is created, its edges serve as one or more boundary conditions for the next injection. Each element can be considered a 'brick' in the larger construct, with the assembly occurring naturally as part of the process. Arbitrary 3D structures can be realized with this approach; we demonstrate here the ability to automate the process using a robotic arm and create a geodesic structure.

LiquiFab structures can be explored both physically and digitally. We provide an open source computational framework, built on Surface Evolver[11] for predicting liquid interfaces in contact with boundary elements and validate it experimentally using 3D scans of produced objects. The software envelope runs Surface Evolver via the Rhino/Grasshopper graphical user interface, without requiring new programming skills.

**Principle of the method**

Consider a volume of liquid polymer injected into an immiscible immersion liquid of equal density. Under such neutral buoyancy conditions, gravity forces are precisely countered by buoyancy such that the only force acting on the liquid is its own surface tension. In the absence of any additional constraints, the liquid would reach a minimum-energy state corresponding to a perfect sphere. This volume can now be transformed into other shapes by bringing it in contact with one or more surfaces that serve as physical constraints or mathematical boundary conditions. The liquid would then take a new shape, corresponding to its new minimum-energy configuration that respects the imposed constraints and its total volume.

Fig. 1a-i provides a schematic illustration of such a transformation. A circular nozzle is used to inject a volume of polymer into a container filled with a mixture of water and glycerol, having a density equal to that of the polymer. The liquid polymer is initially nearly spherical, except for its contact with the injector surface. As the polymer comes into contact with and wets the additional surfaces, it quickly redistributes in space and assumes a new minimum-energy state corresponding to its new geometrical constraints. Addition or aspiration of liquid polymer modifies this minimum-energy state and thus the shape of the surface. The polymer is then cured (either thermally or, for photopolymers, by exposure to UV light), resulting in a solid 3D physical object. In the example here, three circular boundary conditions are used, yet in principle, any number of boundary conditions of any shape can be employed (Supplementary Fig. 5). Importantly, the method is scale-invariant. i.e., once gravity is eliminated, the capillary length becomes infinite, and objects of any size can be created in precisely the same way, as illustrated by the 22 cm tall object in Fig. 1j. The process can be repeated sequentially, with the edges of one object serving as the boundary conditions for the next one, as illustrated in Fig. 1g-i. By automating the process and repeating it, large constructs can be rapidly fabricated. Fig. 1k presents a geodesic dome fabricated using this approach.

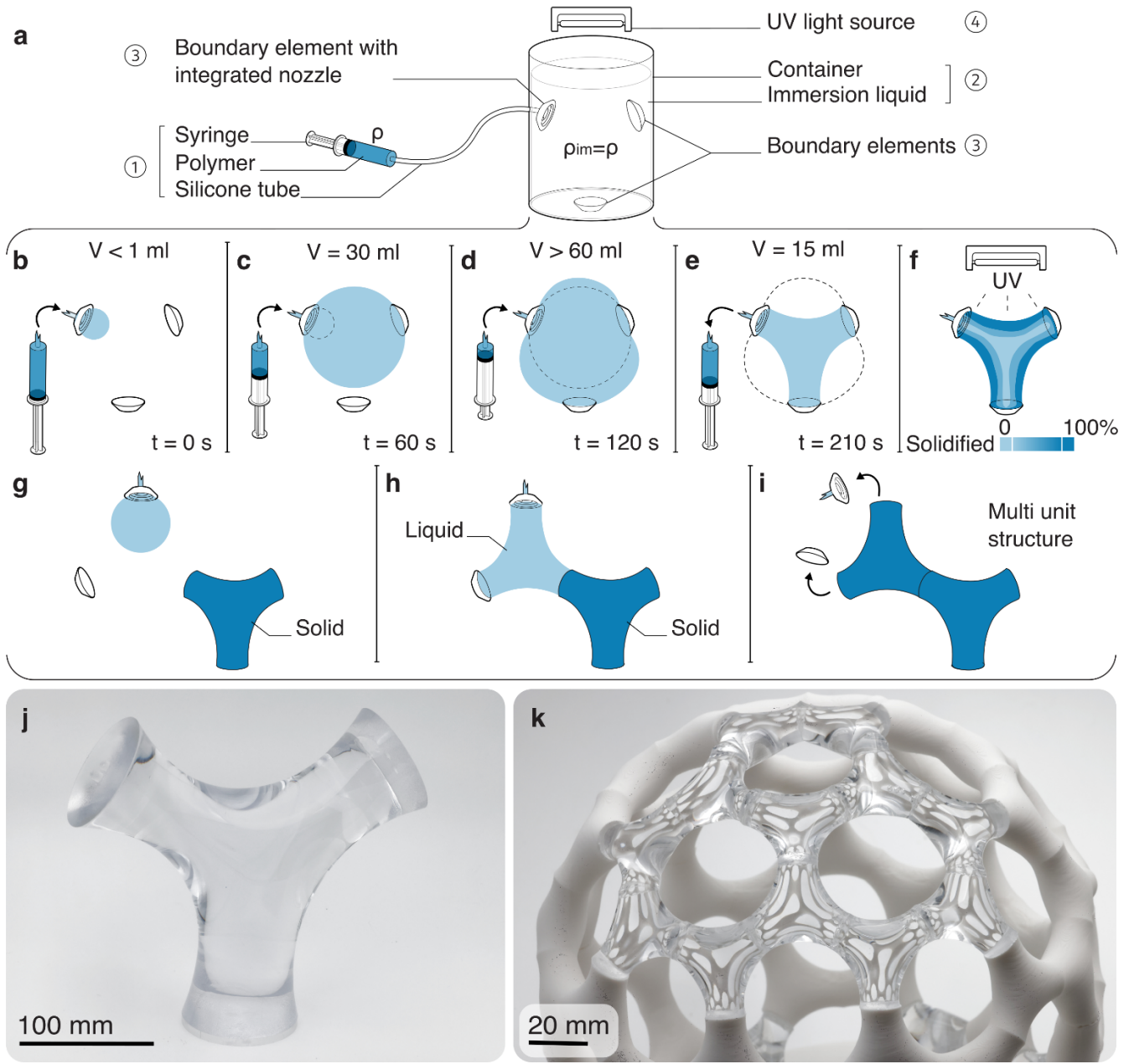

**Fig. 1: The principles of the method. a,** A schematic illustration of the basic elements required for building with liquids – a liquid polymer, a density-matched immersion liquid, a set of boundary conditions, and means to cure the polymer (e.g. thermal, UV). **b–f,** The polymer is injected through one or more boundaries into the immersion liquid, initially forming an approximately spherical shape. As the polymer contacts additional boundaries, it assumes a new minimum-energy state, determined by its volume and the set of geometric constraints. Accordingly, adding or removing polymer changes the shape, which is finally cured to obtain a solid object. **g–i,** This process can be repeated, with the end of the solidified shape serving as a new boundary condition for the subsequent polymer volume. **j,** A solid object, 22 cm tall, built with LiquiFab. **k,** 12 discrete units built sequentially, each serving as the boundary of the other, forming a segment of a geodesic sphere.

**Experimental**

Fig. 2a-d presents our fundamental mode of fabrication (Supplementary video 1), using three equally spaced annular boundary surfaces under a neutral buoyancy environment. Details of the materials and the experimental setup are provided in the methods section. We inject the liquid simultaneously through all three boundaries, forming a nearly spherical 'blob' on each. As more liquid is injected, the blobs grow until they come in contact. Either spontaneously or with slight agitation, the blobs coalesce into a continuous shape corresponding to the minimum-energy state under the prescribed boundary conditions. Importantly, any non-uniformity in the 'blobs' size before coalescing does not affect the final shape, which is determined solely by boundary conditions and the total injected volume. Once the shape settles, we initiate solidification via UV exposure for approximately 10 min. The resulting solid object can then be easily detached from the contact surfaces, which are made of molded silicone, and removed from the immersion liquid. The retrieved object is washed under a water stream to remove any immersion liquid residuals and is then placed to dry in normal room conditions. This example makes use of three boundary conditions with injection through all boundaries. Nevertheless, this approach can readily be implemented using other injection modalities, as shown in Supplementary Fig. 4. It can be extended to a larger number of boundaries, as shown in Supplementary Fig. 5. The method is not limited to UV curable materials and can also be implemented with two-component polymers, as we demonstrate in Supplementary Fig. 6 using polydimethylsiloxane (PDMS).

Fig. 2e-h shows how the method can be adapted for the fabrication of thin-walled structures with internal partitions (Supplementary video 2). As before, we start the process by forming three small 'blobs' (approx. 2 ml in volume each) on the boundaries, which we then 'inflate' by introducing immersion liquid through the boundary ports. As more immersion liquid is injected, the blobs expand until they come in contact and coalesce, forming a hollow structure with three subdivisions. The equal densities of all three liquids (polymer, external immersion liquid, and internal immersion liquid) ensure neutral buoyancy throughout the system, leaving surface tension as the only force governing the dynamics of the liquid interfaces. One should, however, note that the outer envelope of the structure differs from that of the full polymer depicted in Fig. 2a-d. This is because the thin partitions undergo an extremely slow draining process, and the resulting solidified object therefore captures a time point in a dynamic process rather than in equilibrium.

Fig. 2i-l depicts a sequence of liquid polymer transformations due to the linear movement of a single boundary (Supplementary video 3). As the boundary moves toward the center of the form, the liquid adapts and settles into a new minimum-energy state. Surprisingly, a stable shape is maintained even when the angle between the liquid and the surface exceeds 180 degrees (Fig. 2k), resulting in a Napoleonic Bicorn shape (Fig. 2l).

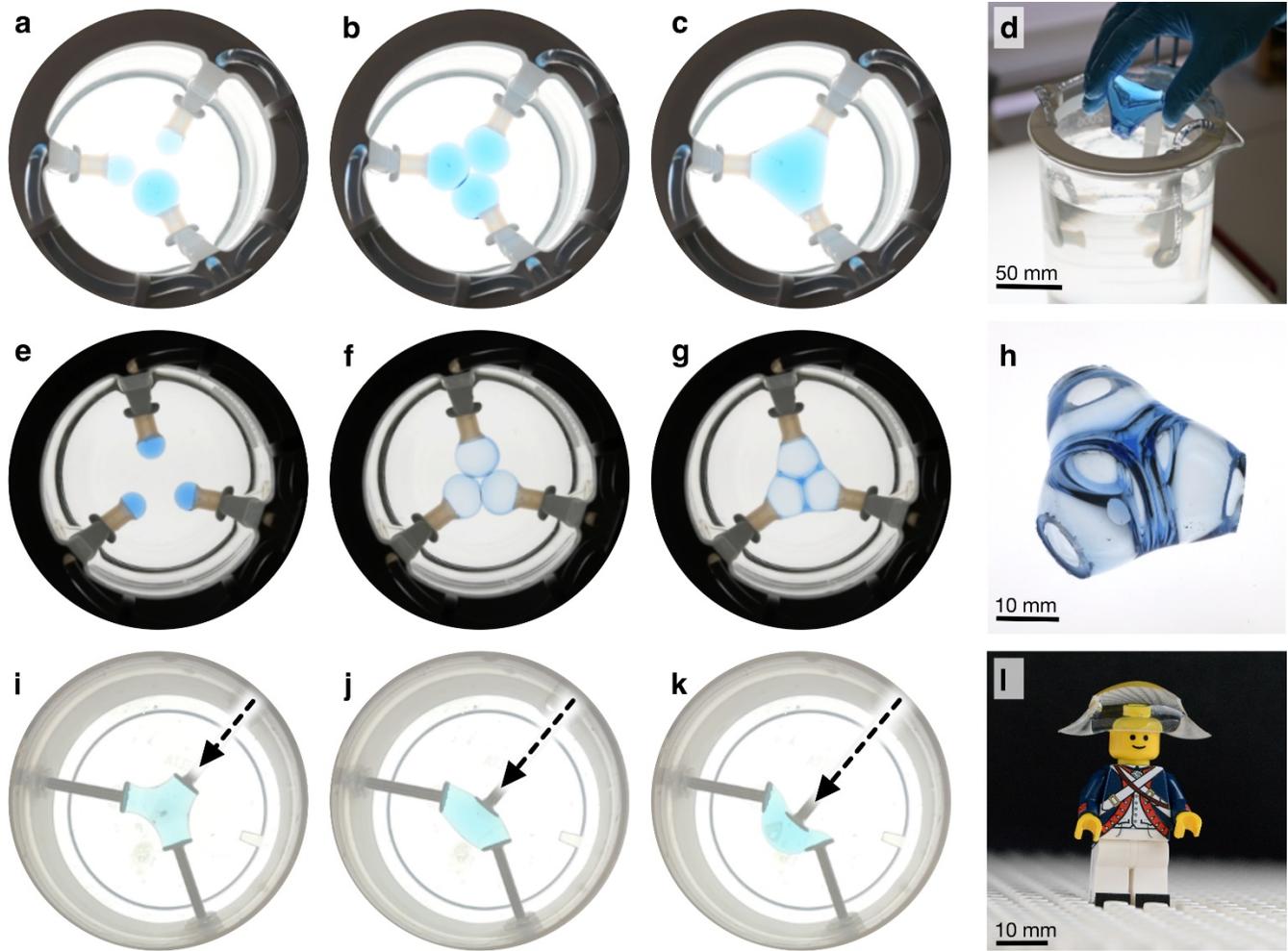

**Fig 2: Modes of fabrication. a-d, Fabrication of a homogenous object**. Polymer is simultaneously injected into the immersion liquid from three boundaries, initially forming standalone blobs. Upon further polymer injection, the blobs come into contact and coalesce into a single continuous body. The amount of total polymer inflow controls the final shape of the body. Following a ten-minute cure under UV, the solid object is complete. **e-h, Fabrication of a hollow multi-chamber shape.** A small liquid polymer blob is initially placed on each boundary. Immersion liquid is injected into the polymer blobs, forming thin-walled spheres. Upon contact, the spheres coalesce and transform into a unified body with internal partitions. After polymerization, the inflation liquid is drained, resulting in an object with internal cavities divided by internal walls. **i-l, Fabrication by moving boundaries.** For a polymer volume in contact with all its boundaries, the displacement of boundary elements forces it to adjust to a new minimum-energy state, while the total polymer volume in the system remains constant throughout the process. This allows the fabrication of shapes that are not achievable with static boundaries.

## Computational framework

Surface Evolver is a powerful computational tool developed by Ken Brakke[11] for simulation of the minimum-energy state of liquid volumes and surfaces. Surface Evolver's robust mathematical framework enables the incorporation of a variety of forces and boundary conditions. It has been used in a wide range of applications, such as the behavior of liquid rocket fuel in low gravity, the shape of molten solder in microcircuits, and the formation of triply periodic minimal surfaces. It is thus a natural choice for simulating LiquiFab structures.

To enable computation of complex structures and rapid iterations on design ideas, as well as to make the tool accessible to the design community, we developed a computational envelope in Grasshopper/Rhino. This Grasshopper-Surface Evolver (GSE) environment facilitates simple definition of boundary conditions and liquid volumes, and allows to run the Surface Evolver computational core through a graphical user-friendly environment (see Methods section).

As shown in Fig. 3a, the user defines the boundary conditions by creating an arbitrary closed surface that represents the boundaries of the injected liquid. We designed GSE to interpret any holes in this envelope as fixed boundary conditions to which the liquid will be pinned. The precise shape of this surface is of no consequence and can be defined as a simplistic block. The GSE then creates a Surface Evolver command language script that includes the mesh itself, a convergence algorithm to settle the fluid into its final form, and a remeshing algorithm based on Bostch et al.[12] applied during the convergence process. Surface Evolver is executed as a subprocess of the Grasshopper script. It outputs its converged result to a file, which is then automatically reloaded into Grasshopper, enabling further handling in the Grasshopper/Rhino environment. Fig. 3b-d show examples of such solutions obtained using the boundary conditions depicted in Fig. 3a. In perfect neutral buoyancy, an increase in the liquid volume results in the inflation of the original shape (Fig. 3b-c), while a slight deviation from neutral buoyancy yields out-of-plane deformations of the original shape (Fig. 3d). The run time of the simulation depends on the geometry and the desired surface resolution but is on the order of few seconds for the examples we show here, making it a convenient tool for iterative design.

To validate the numerical code, we compare physical and computational objects produced under the same conditions. Fig. 3e presents an image of a physical object we produced using the boundary conditions and volume of the computed object in Fig. 3c. By scanning the object with a 3D scanner, we obtain its surface topography in digital form, which we can then compare to the computed topography. Fig. 3f presents such a comparison as a 3D colormap of the surface difference between the two (Methods section). The differences are sub-millimetric across the entire object, with a maximum of 0.5 mm and an RMS of 0.2 mm, representing deviations of 1.2% and 0.5% of the characteristic body length, respectively. These deviations are very reasonable and even expected, given polymer shrinkage, immersion liquid density errors, and scanning accuracy, and thus provide confidence in the validity of the computational tool.

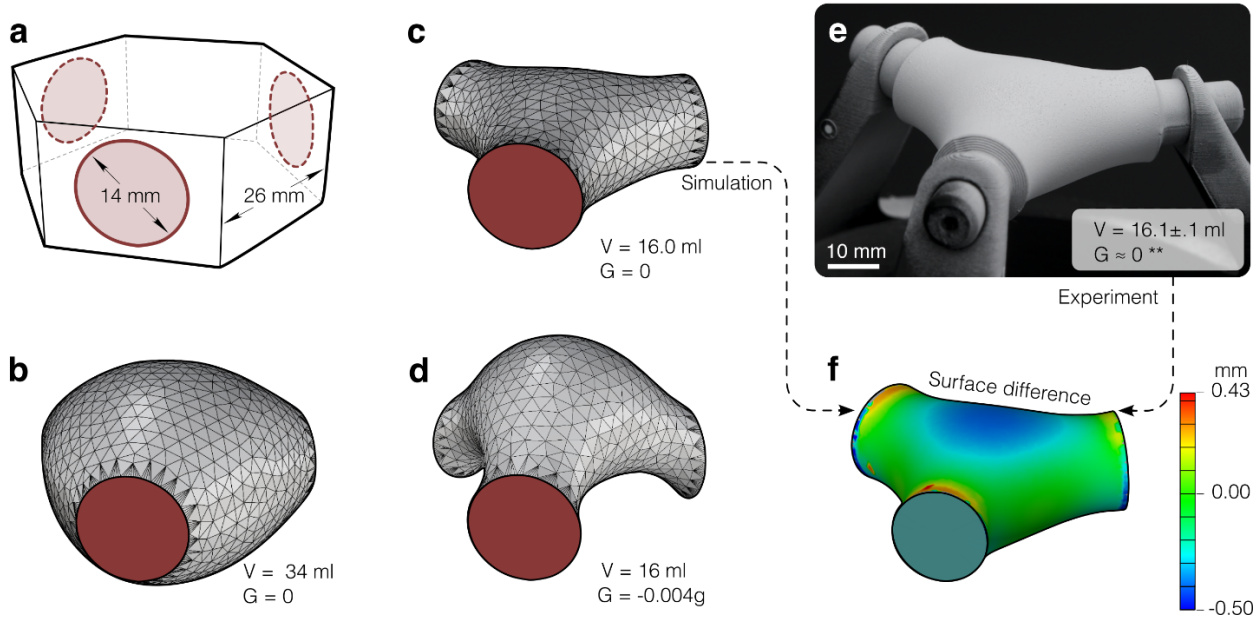

**Fig 3: Demonstration and validation of the computational tool. a-d,** Simulation of LiquiFab components using the Grasshopper-Surface Evolver tool. **a,** The simulation receives as input the fixed boundaries (depicted as red circles) within an arbitrary closed surface envelope (here, a simplified polyhedron), a target volume, and relative gravity. **b,** The resulting shape for a volume of 34 ml under zero gravity. **c,** The resulting shape for the same boundaries with a volume of 16 ml under zero gravity value. **d,** The resulting shape for the same boundaries and volume as in **c**, but with a residual buoyancy ( G = -0.004g). **e,** A LiquiFab physical object, fabricated using a neutral buoyancy environment (G = 0) and with a nominal volume of 16 ml. **f,** A comparison between the computational and fabricated model. The color map shows the difference between the objects, with an RMS deviation of 200 micrometers.

# Sequential Construction from LiquiBricks

To fabricate more complex structures and assemblies, we start by segmenting a desired surface into planar facets - a standard technique in computational geometry and graphics (e.g., via variational shape approximation[13]). Each facet is then replaced by a set of virtual boundary surfaces whose geometrical centers coincide with the centers or the edges, and whose normal is perpendicular both to the edge and to the average normal of the facets connected by the edge. For simplicity, we use disk-shaped boundaries with a fixed diameter, though in principle both the shape and size of the boundaries can vary within an object. Each set of bounding surfaces defines a LiquiBrick – an object that would be formed upon injection of a liquid volume connecting those boundaries. Using our GSE code, we create a digital prediction for the entire network and use it to determine the volume assigned to the LiquiBrick (which may be uniform or vary across the geometry). As an illustrative example, in Fig. 4a-c, we demonstrate this concept using a geodesic dome (an icosahedral sphere with a frequency of 2), originally popularized by architect Buckminster Fuller.[14] (Supplementary video 4).

Fig. 4d presents the system we developed to implement such structures. The system is built around a six-degree-of-freedom robotic arm, computer-controlled for precise positioning. As its end effector, the arm holds a metal pipe fitted with a silicone nozzle. Liquid polymer is delivered to the nozzle by a computer-controlled syringe pump through a silicone line. The nozzle is submerged within an immersion liquid contained in a glass vessel placed on a transparent plate. The plate transfers light produced by UV LEDs from below, enabling in situ polymerization of the injected volumes. For uniform curing of completed objects, the container is enclosed within a polymerization sphere consisting of a base and a detachable lid. When closed, this chamber provides isotropic UV illumination over the entire construct.

The fabrication process is outlined in Fig. 4e-g. Given two existing boundary surfaces, the robot positions the nozzle near one of the boundaries and injects a small initial volume of polymer that wets that boundary (Fig. 4e), and then retracts from it. This is then repeated for the second boundary. Next, the robot positions itself so that the nozzle edge lies between the two boundaries and injects a prescribed volume of polymer, which coalesces with the polymer deposited on the two wetted boundaries, forming a continuous volume of liquid connecting the two surfaces and the nozzle (Fig. 4f). Finally, the nozzle moves to its last position thereby shaping the polymer into the desired form (Fig 4g). The UV light is turned on for a short time (24 s) to partially polymerize the LiquiBrick. Illuminating only for a short time ensures that only the outer shell of the object is solidified, while its inner volume remains liquid. The nozzle then retracts, forming a new stand-alone object. The process then repeats with the next LiquiBrick. Keeping the core of the object in liquid form ensures facile coalescence with the liquid injection of the next object. Once the entire structure is formed, it is exposed to UV for a longer duration until it is fully cured (Fig. 4h). (Supplementary video 4).

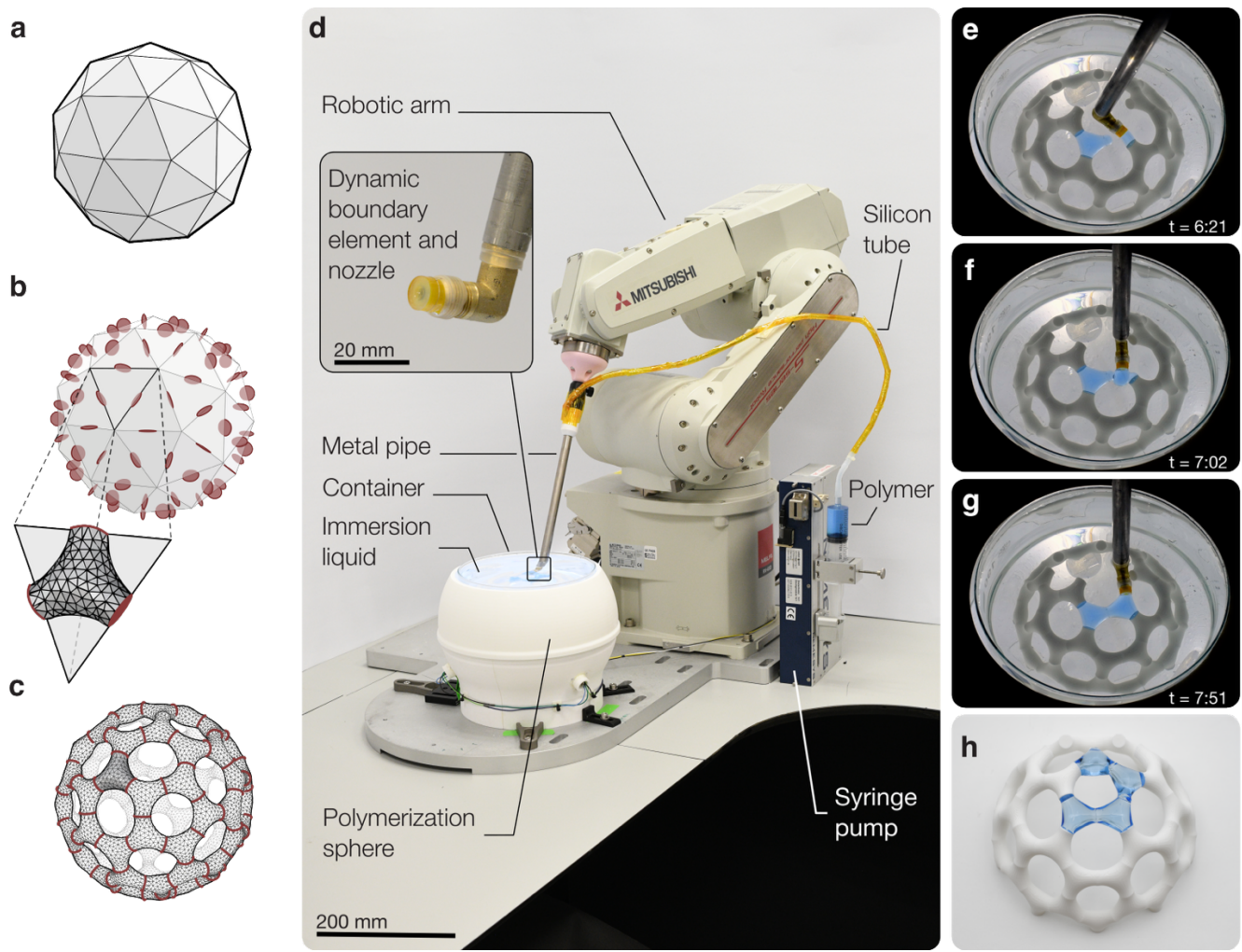

**Fig 4: Computational design and robotic fabrication of a geodesic construct. a-c,** A geodesic dome design based on a 2v-icosahedron. **a,** the 2v-icosahedron introduced by architect Buckminster Fuller. **b,** At each edge's midpoint we define a circular boundary surface with a normal that is perpendicular both to the edge and to the average normal of the facets connected by the edge. Each face is thus replaced by a three-dimensional object, as illustrated by the simulation. **c,** Simulation of the geodesic dome obtained using the edge boundaries. **d,** The automated fabrication system uses a 6 DOF robotic arm to control the location of a boundary element within an immersion liquid, and creates an object through a sequence of polymer injections and polymerizations. **e-g,** Time-lapse images showing the main fabrication steps. **e,** Polymer wetting of an existing boundary. **f,** Retraction and further injection to bring adjacent boundaries into liquid contact. **g,** Repositioning while further injecting to create the element's final shape. UV light is then turned on for polymerization (not shown). **h,** The fully polymerized construct.

**Discussion and Conclusions**

We demonstrated a fabrication method that enables the creation of three-dimensional structures directly from liquid polymers, shaped by boundary conditions in weightlessness. The approach, which we termed LiquiFab, extends the concept of form-finding[1] into a practical manufacturing process. It is scale-invariant, zero-waste, and capable of producing continuous, structurally robust volumes within minutes. It is compatible with a wide range of materials and can produce both soft and rigid objects.

LiquiFab defines a new class of manufacturing methods, differing fundamentally from additive processes such as fused deposition modeling (FDM)[9], rapid liquid printing (RLP)[15], or stereolithography (SLA)[10]. In all of these, the printing head or light pattern must effectively visit every voxel in space to define the final geometry. These approaches provide exceptional geometric freedom but are intrinsically non-scalable in time, as printing duration grows directly with the volume of the object. A fundamentally different strategy is introduced by volumetric additive manufacturing via tomographic reconstruction (VAM)[16], in which the entire object is formed simultaneously rather than through sequential voxel addressing. However, VAM relies on optical penetration through the photopolymer, imposing physical limits on object size and material choice.[17] Moreover, only a fraction of the resin volume contributes to the final solid, as the surrounding polymer is typically discarded after exposure.[16]

LiquiFab cannot provide the full geometric freedom offered by 3D printing methods [9,10,16], but within its accessible geometric space, it provides unprecedented advantages. The formation of 3D objects is governed solely by interfacial physics and is independent of light propagation or voxel-by-voxel motion. This is particularly advantageous for large-scale constructions, where conventional printing would become prohibitively slow, and where the overall form and strength of the structure take priority over fine geometrical details. Because the formed body is fluidically continuous, it naturally yields homogenous and smooth structures, without the characteristic layers of 3D printing.

The practical scale of a structure fabricated under simulated weightlessness is limited by two factors: the maximum size is bounded by the physical dimensions of the immersion liquid container, and by imperfect density matching, which leads to a finite capillary length. In space, LiquiFab is entirely free of these limitations, as an immersion liquid is no longer necessary. Therefore, the process is not confined to a manufacturing envelope, and structures could, in principle, grow to any size. The combination of zero-waste operation, scale invariance, and insensitivity to gravity makes LiquiFab a promising candidate for in-situ fabrication of large structures in space. Several studies, including Yerazunis et al[18] and Crowe et al[19], have already developed space-grade formulations that can withstand deep vacuum, radiation, and atomic oxygen, and could thus be excellent candidates for LiquiFab in space.

Our present demonstration used a single nozzle as the end effector. With one injector, the constructed geometry is always converging—two existing boundary conditions merge into a single new one. Future implementations may employ dual or multiple injectors, such that each boundary could lead to two or more new ones. This would enable the creation of diverging geometries, such as tree-like branches or networked structures. Another useful enhancement would be to allow the injector to rotate freely on a ball bearing, enabling injection from orientations currently inaccessible due to geometric constraints of the robotic arm.

As demonstrated in this work, Surface Evolver [11] is useful for prediction of the shape of individual building blocks, as well as for the design of complex structures such as the geodesic dome. However, it is based on energy minimization and is designed to compute liquid volumes at steady state. As discussed in the experimental section, the hollow partitioned structures (Fig. 2e-h) are inherently not at steady state; their thin partitions represent an extremely slow-draining process that is effectively frozen by polymerization. Predicting such transient structures would therefore require a fundamentally different, time-dependent computational approach capable of resolving the dynamics of fluid evolution.

# Methods

## Materials

To create a simulated zero-gravity environment, we filled a glass container with an immersion liquid consisting of Deionized (DI) water and glycerol (Bio-Lab, Israel; Cat No. 000712050100). We tuned the density of this solution to match the density of the specific liquid polymer under test. To ensure precise density matching, we monitored the solution's concentration using a refractometer (ATC, USA; Range: 0–32% Brix).

We prepared two types of polymers: a single-component UV-cured resin and a two-component silicone elastomer. Following preparation, we degassed both polymer types under a negative pressure of 0.7 to 0.8 atm to remove trapped air bubbles. UV-cured resin: We used Vida Rosa UV resin, optionally colored with a translucent resin dye (Miraclekoo). We adjusted the immersion liquid for this material to a Brix value between 25% and 27%, depending on the concentration of the colorant dye (Fig. 1j-k, Fig. 2, Fig. 3e, Fig. 4e-h, and supplementary Fig. 5a). PDMS Elastomer: As a two-component control, we utilized the SYLGARD 184 Silicone Elastomer Kit. We adjusted the immersion liquid for the PDMS experiments to a Brix value of $9.5 \pm 0.2\%$ (Supplementary Fig. 5b).

## Experimental systems

**Basic system.** Supplementary Fig. 1 presents the experimental setup used in Fig. 2. The system is based on a 2000 ml glass (borosilicate 3.3) cylindrical container with a diameter of 13 cm. Resting on top of the container is a 'Fixator' – a printed (PLA; Bambu Lab, X-1 Carbon) element that consists of a circular ring from which multiple arms extend into the container. The arms hold the bounding surfaces, which were full when they served only as geometrical constraints and hollow when they were also used as sources for polymer injection. The boundaries were either printed as an integral part of the fixator or made from molded silicon (Specialty Resin & Chemical; Cast-a-mold platinum food-grade silicone rubber, Hardness 20 shore A) when easy detachment of the final object was desired. When serving as nozzles for liquid injection, the bounding elements were connected via silicone tubing to a syringe, which was operated either manually or with a syringe pump (Cole-Parmer, USA; Cat No 74900-15).

The container was placed on three poles within a 51 cm diameter plaster polymerization sphere, internally coated with Barium Sulfate ($BaSO_4$) powder (CS-chemicals, Israel; CAS No. 7727-43-7), such that the 3D polymer shape is roughly centered. The polymerization sphere contained three 900 mW UV 365 nm LED sources (Luminus Devices; P/No SST-10-UV-A130-G365-00) equally spaced along its equator. Three circular optical baffles (coated on both sides with $BaSO_4$) with a 50 mm diameter were located 40 mm in front of each LED. The entire setup was constructed on a 60 cm by 60 cm honeycomb breadboard optical table (Thorlabs, USA; P/NoB6060A) to suppress vibrations.

**Robotic System.** Supplementary Fig. 2 depicts the components of the robotic system shown in Fig. 4. At the heart of the system is a silicone nozzle that serves as both a boundary condition and the injection point for the polymer. The nozzle was mounted on a brass elbow, which was attached on its other end to the tip of a 22 cm hollow stainless-steel pipe. The metal pipe was secured to the endplate of a six-degree-of-freedom robotic arm (Mitsubishi, Japan; MELFA RV-6SD) using a custom 3D-printed adapter coupled with standard mechanical fittings.

Liquid polymer was supplied to the nozzle through a silicone tube that led from a 60 cc syringe mounted on a syringe pump (Cetoni, Germany; Nemesys), through the pipe, and into the brass elbow. Any segment of the polymer route that could be exposed to stray light or the polymerization light was wrapped with Kapton tape (DuPont, USA; Kapton) to prevent premature polymerization.

The polymerization sphere was composed of a base that housed the water container and a slitted lid that allowed the nozzle to remain submerged even when the lid was closed. During the robotic operation, the sphere was kept open to allow free motion of the arm, and the lid was placed only during polymerization steps. The base included a clear glass plate fixed at approximately 11 cm from the bottom. Three UV LEDs (Luminus Devices; P/No SST-10-UV-A130-G365-00) covered with 35 mm optical baffles were located at the bottom of the sphere such that they illuminated the volume through the glass plate. All structural components of the sphere - the base, lid, and baffles – were 3D-printed in white PLA (Bambu Lab X-1 Carbon) and coated internally with $BaSO_4$ (CS-Chemicals, Israel; CAS No 7727-43-7) to ensure uniform diffuse reflection.

The immersion liquid was contained in a 2000 ml borosilicate (3.3) cylindrical vessel, 19 cm in diameter, which was affixed to the glass plate using double-sided mounting tape (Gorilla Glue Company, USA; Gorilla Tough and Clear).

The robot was programmed with Mitsubishi's 'MELFA-Basic V' programming language and operated using its dedicated software (Mitsubishi, Japan; RT toolbox 3). The syringe pump was operated via the Cetoni software (Cetoni, Germany; neMESYS UserInterface v 2.6.0.4).

## 3D scanning and surface comparison

Scanning was performed using a structured light scanner (Zeiss, Germany; GOM Scan 1 (200)) with the object placed on a rotating table (Zeiss, Germany; GOM ROT 350). The Zeiss Inspect software automatically stitches the obtained images to produce the three-dimensional digital representation. Because the LiquiFab objects are highly transparent, they were coated with a single layer of masking spray (AESUB, Germany; AESUB orange) estimated by the manufacturer to have a thickness of 5-6 microns. Comparison between the scanned geometries and the computationally predicted ones was also performed using the Zeiss Inspect software using the following parameters: 'Prealignment' - ON, 'Compute additional best-fit' – ON, 'Create Surface Comparison On CAD' with 'Max. Distance' = 0.50 mm, 'Collect local deviation peaks' – ON, 'Max. deviation of normal' = 60°, 'Max. opening angle' = 30°.

## Code availability

The Grasshopper Surface Evolver (GSE) tool is freely available at https://github.com/Amir-Mann/surface_evolver_grasshopper/.

# Acknowledgements

We thank Technion's Crown-Vanguard Fund for its financial support of the project. We thank Alexey Razin for assistance with the electronic equipment, Roni Hillel for aid with the Grasshopper software, Moriya Hadad for assistance with the computational code, Haim Singer for assistance with video captures, and Mor Elgarisi for hands-on assistance in the lab.

# Author contributions

E.H., V.F., A.S., and M.B. conceived the research; E.H., A.S., and M.B. designed the experiments; E.H. performed the experiments and the analysis; K.S. and Y.M. assisted with the experiments; A.M. led the algorithmic and computational work under the guidance of E.H. and M.B.; E.H., K.S., and Y.M. prepared the figures; E.H., A.S, and M.B. wrote the manuscript; All authors reviewed and commented on the manuscript.

# Competing interests

The authors declare no competing interests.

# Materials & Correspondence

Aaron Sprecher, asprecher@technion.ac.il
Moran Bercovici, moran.bercovici@mat.ethz.ch

# Supplementary Information

# LiquiFab – Building with liquids in weightlessness

Erez Hochman[1,2], Aaron Sprecher[2,*], Kateryna Suzina[2], Amir Mann[3], Yuval Mihalovich[2], Valeri Frumkin[4], Moran Bercovici[1,5*]

[1] Faculty of Mechanical Engineering, Technion – Israel Institute of Technology, Haifa, Israel
[2] Faculty of Architecture, Technion – Israel Institute of Technology, Haifa, Israel
[3] Faculty of Computer Science, Technion – Israel Institute of Technology, Haifa, Israel
[4] Boston University, Boston, MA, USA
[5] Department of Materials, ETH Zürich, Zürich, Switzerland

* Corresponding authors: asprecher@technion.ac.il, moran.bercovici@mat.ethz.ch

## Supplementary Videos

Supplementary Video 1. Fabrication sequence and experimental system

Supplementary Video 2. Transformation of liquid polymer into a three-chamber shell structure

Supplementary Video 3. Transformation of liquid polymer with moving boundaries

Supplementary Video 4. Design framework and robotic construction of geodesic elements

# Supplementary figures

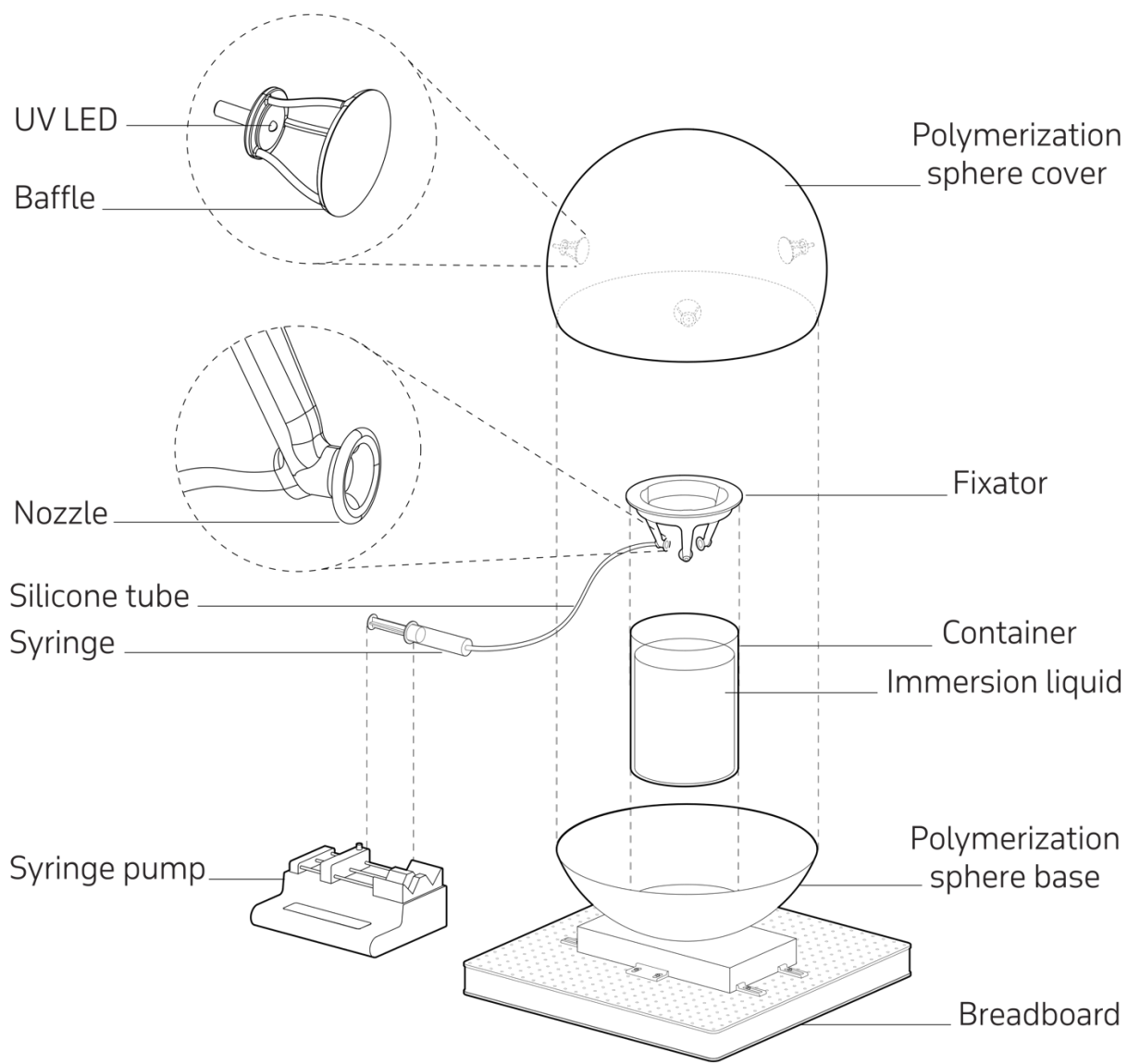

**Supplementary Fig. 1: the basic experimental system – exploded view.** The polymeric objects are created inside the container containing an immersion liquid. The polymer injection is driven by a syringe pump, which is connected via a silicone tube to one or more nozzles submerged in the container. The nozzles are held in space through a fixator that rests on the upper edge of the container. To ensure uniform polymerization, the container is enclosed within a 'polymerization sphere' containing multiple UV LEDs. Each LED is covered by a baffle so that the polymer is not illuminated directly by the light sources, but instead receives light that the sphere's inner surface has diffused. The 'polymerization sphere' is mounted on a breadboard to filter vibrations from the environment.

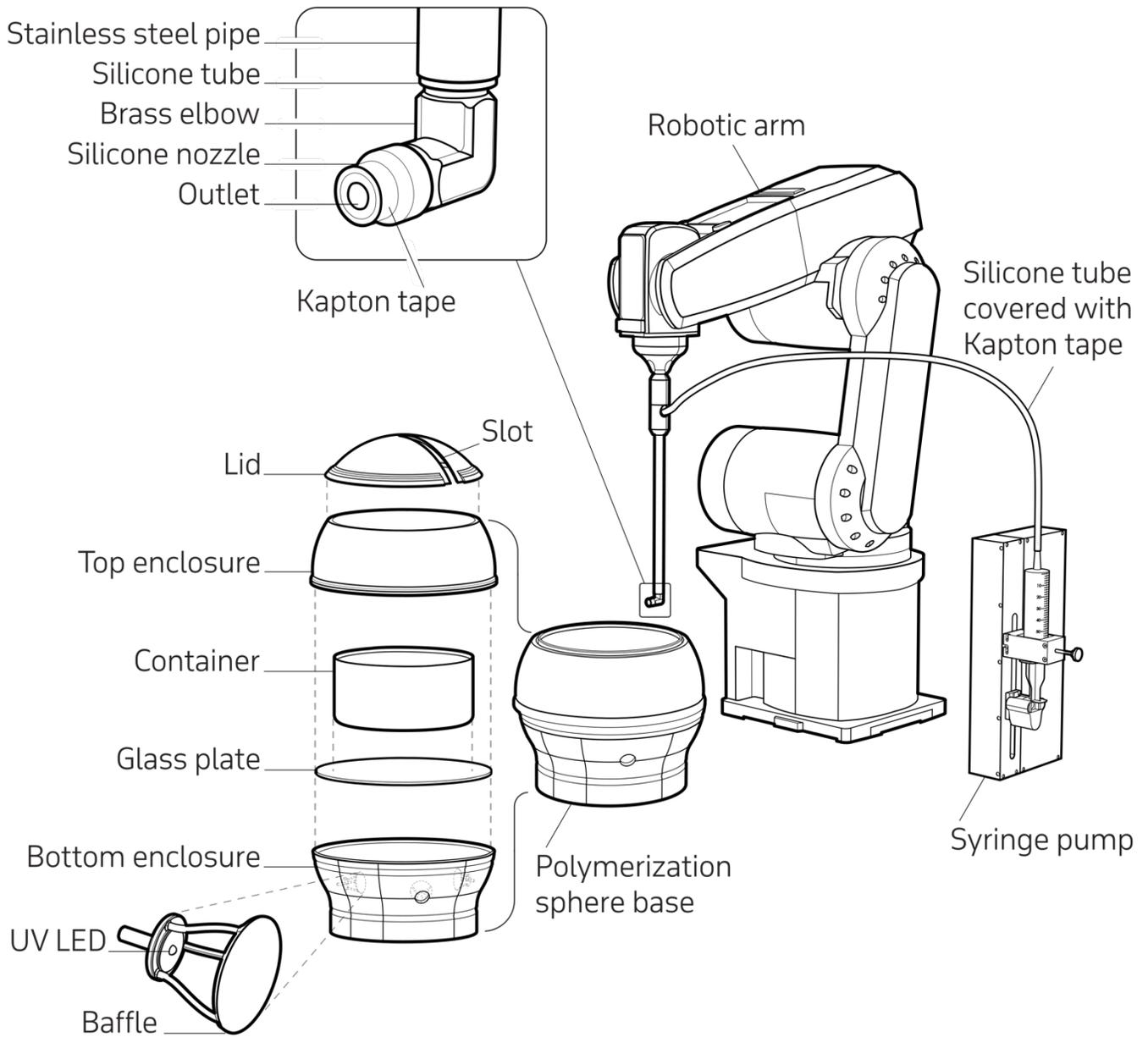

**Supplementary Fig. 2: robotic experimental system.** The robotic system is based on a 6DOF robot that controls the location of an injection nozzle in space, operating in sync with a syringe pump that supplies the polymer to the nozzle. The nozzle is submerged in an immersion liquid container, fixed to the top of a transparent glass plate within a polymerization sphere. Under the plate are three baffled UV LEDs that serve as light sources for the polymerization phases of the sequence. The slitted lid allows the nozzle to remain in place even during polymerization. Any exposed regions of the polymer's flow path are covered with Kapton tape to prevent stray light from initiating premature photopolymerization.

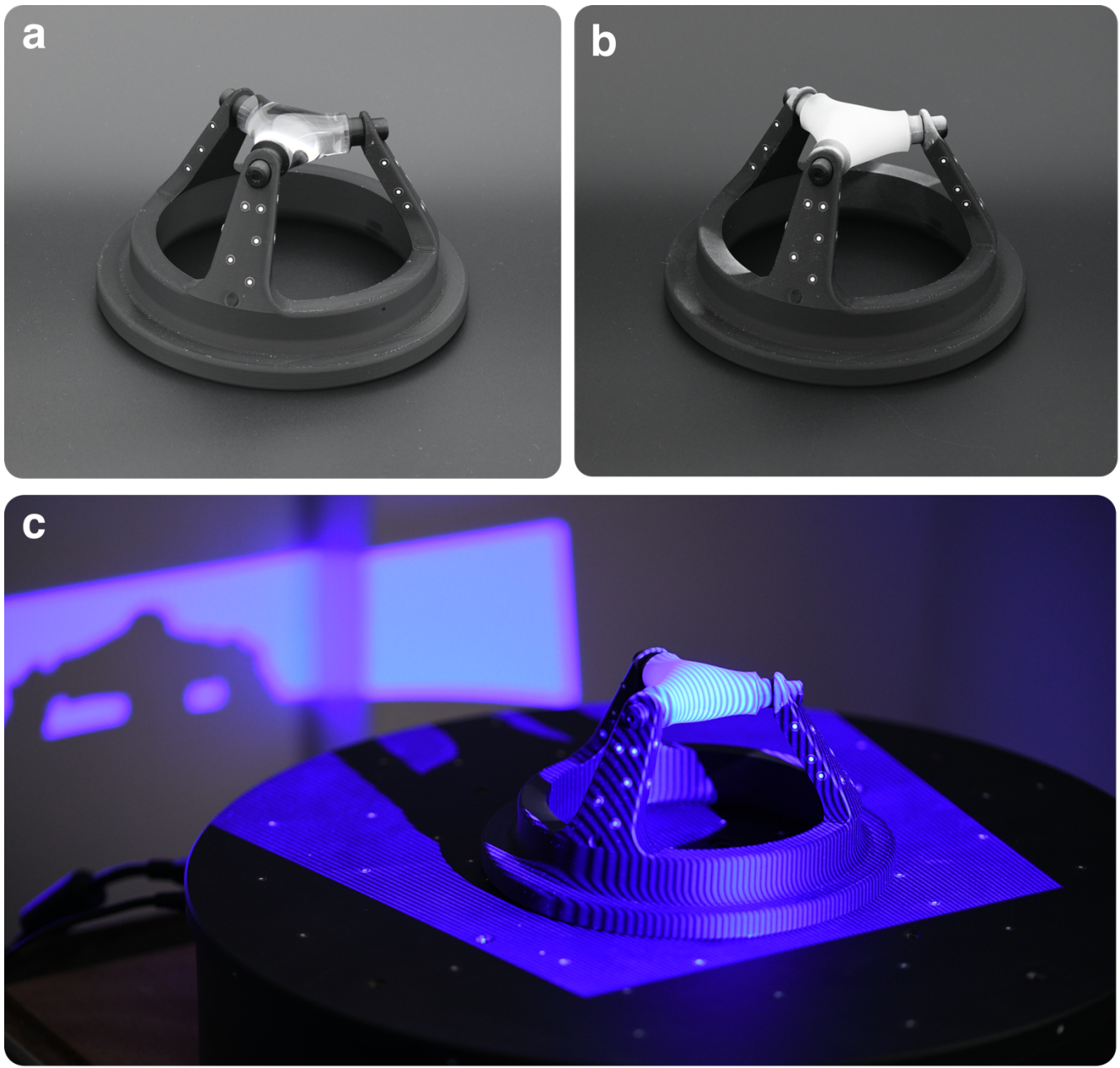

**Supplementary Fig. 3: LiquiFab sample preparation and 3D scanning. a,** A transparent solidified sample, still attached to its 3D printed fixator. The white markers attached to the fixator serve as position indicators for the 3D stitching algorithm. **b,** The same sample from **a** covered with a white opaque layer of scanning spray. **c,** The sample from **b** on a rotating table, during the scan.

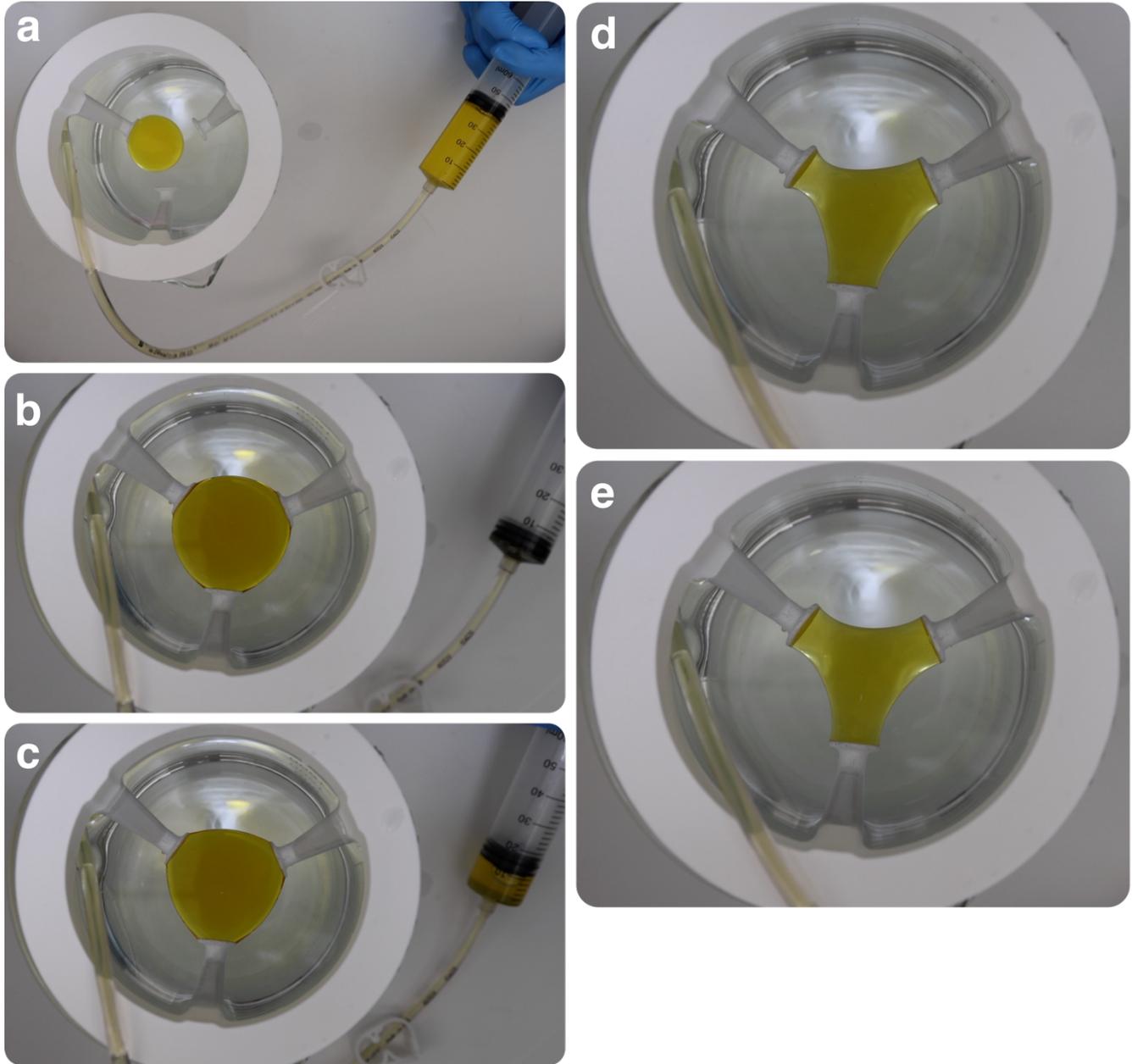

**Supplementary Fig. 4: Top view of a time-lapse of a fabrication process using a single injection point**. **a,** We inject liquid polymer into the immersion liquid through one of the bounding surfaces, initially creating a spherical shape. **b,** As more polymer is injected, the sphere grows until it reaches the other boundaries and wets them. **c,** Once all three boundaries are in contact with the liquid polymer, we aspirate some of the polymer back into the syringe. **d,** As polymer is removed, capillary forces act against viscosity to adjust the shape of the object. For this polymer and scale, the process takes several seconds. **e,** The shape settles to a new minimum energy state.

Page **5** of **7**

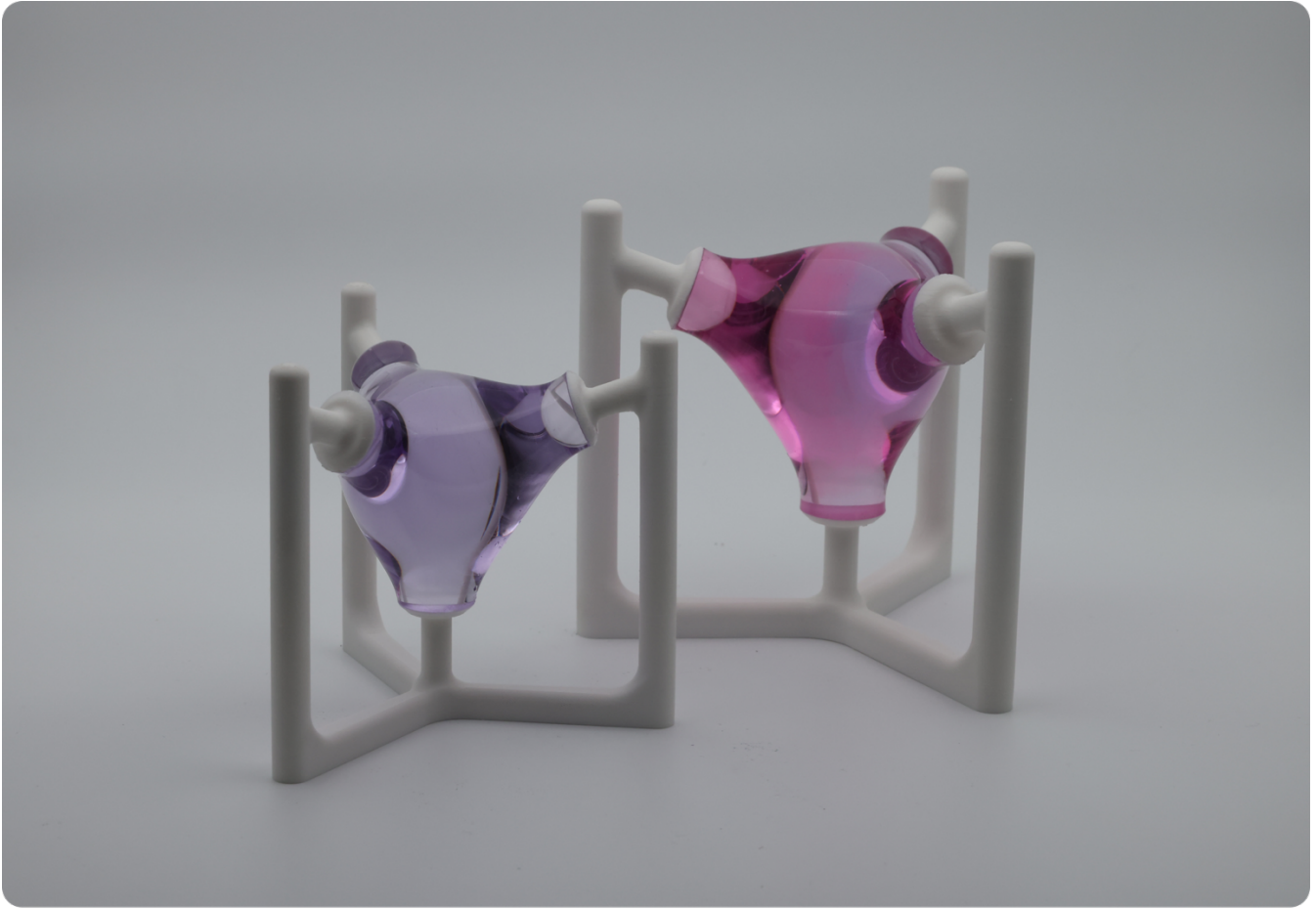

**Supplementary Fig. 5: LiquiFab objects with four bounding surfaces.** Two colored solid objects of different sizes made of UV curable polymer, using four equally spaced bounding surfaces, whose normals meet at a single point at the center of the object.

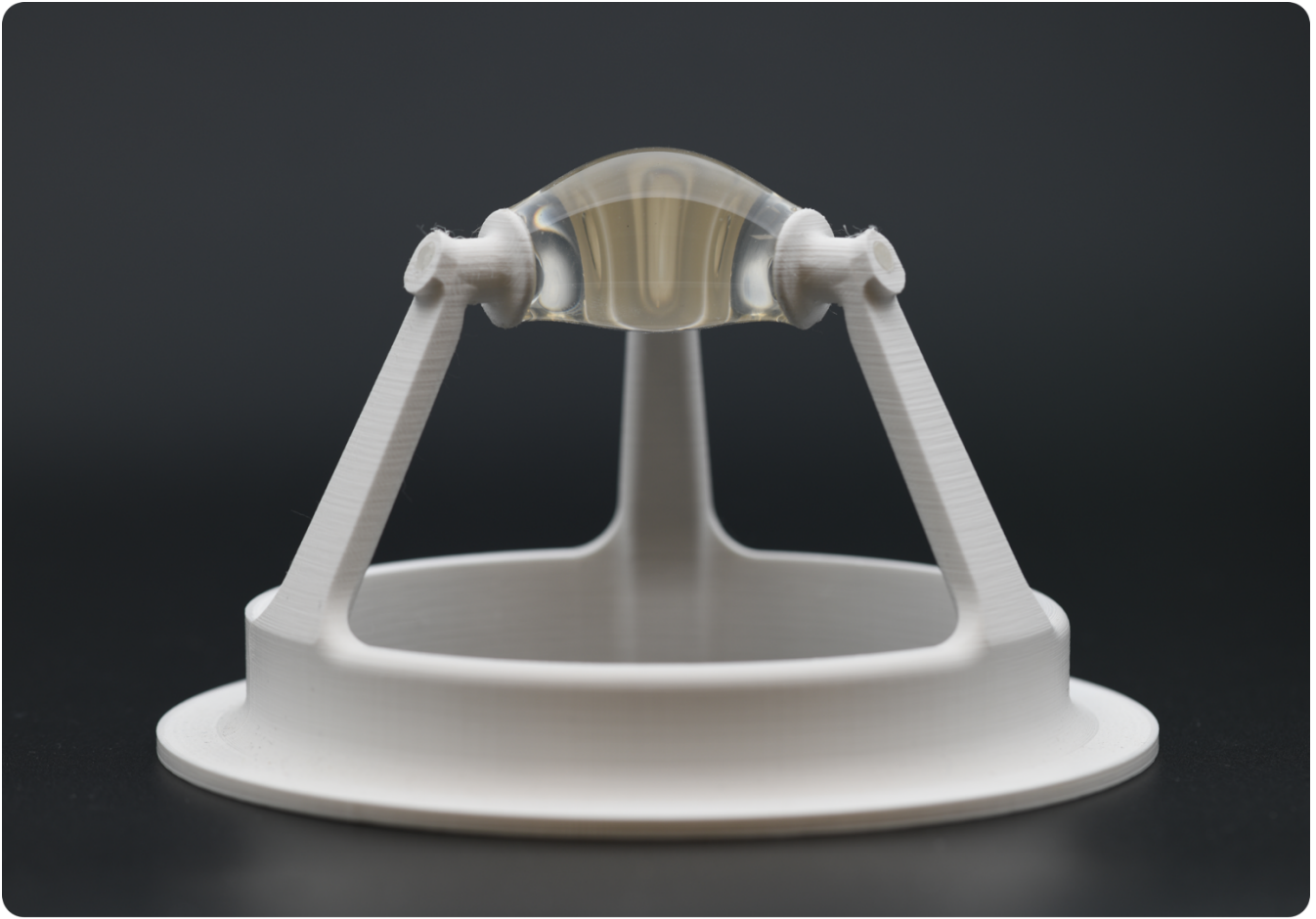

**Supplementary Fig. 6: LiquiFab object made of PDMS.** Most of the objects presented in this work were made from a photocurable material. However, the method works equally well for thermally cured material, such as polydimethylsiloxane (PDMS).